\documentclass[pre,aps,floats,superscriptaddress,floatfix,twocolumn,longbibliography]{revtex4-1}
\usepackage{amssymb,amsmath}
\usepackage{amsmath,amssymb}
\usepackage{enumitem,kantlipsum}
\usepackage{lipsum}

\usepackage{gensymb}
\usepackage{graphicx}
\usepackage{psfrag}
\usepackage{color}
\usepackage{dcolumn}
\usepackage{bm}
\usepackage[normalem]{ulem}
\usepackage[absolute]{textpos}
\usepackage{hyperref}
\hypersetup{
  colorlinks=true,
  citecolor=magenta,
  linkcolor=cyan,
}

\begin{document}
\title{
Logarithmic or algebraic: roughening of an active Kardar-Parisi-Zhang surface\\
}
\author{Debayan Jana}\email{debayanjana96@gmail.com}
\affiliation{Theory Division, Saha Institute of Nuclear Physics, A CI of Homi Bhabha National Institute,  1/AF Bidhannagar, Calcutta 700064,West Bengal, India}
\author{Astik Haldar}\email{astik.haldar@gmail.com}
\affiliation{Department of Theoretical Physics $\&$ Center for Biophysics,
Saarland University, 66123 Saarbr\"ucken, Germany}
\author{Abhik Basu}\email{abhik.123@gmail.com, abhik.basu@saha.ac.in}
\affiliation{Theory Division, Saha Institute of Nuclear Physics, A CI of Homi Bhabha National Institute,  1/AF Bidhannagar, Calcutta 700064,West Bengal, India}

\begin{abstract}

The Kardar-Parisi-Zhang (KPZ) equation sets the universality class for growing and roughening of nonequilibrium surfaces  without any conservation law  and nonlocal effects. We argue here that the KPZ equation can be generalized by including a symmetry-permitted nonlocal nonlinear term of active origin that is of the same order as the one included in the KPZ equation. Including this term, the 2D active KPZ equation is stable in some parameter regimes, in which  the interface conformation fluctuations  exhibit sublogarithmic or superlogarithmic roughness, with nonuniversal exponents, giving positional generalised quasi-long-ranged order. For other parameter choices, the model is unstable, suggesting a perturbatively inaccessible algebraically rough interface or positional short-ranged order. Our model should serve as a paradigmatic nonlocal growth equation.

\end{abstract}

\maketitle


{ The Kardar-Parisi-Zhang (KPZ) equation~\cite{kpz,kpz1,stanley}  for growing nonequilibrium surfaces displays a nonequilibrium roughening transition between a smooth phase, whose long wavelength scaling properties are identical to an Edward-Wilkinson (EW) surface~\cite{ew}, to a perturbatively inaccessible rough surface~\cite{stanley,natter} when $d>d_c=2$, its lower critical dimension. Importantly, the local normal velocity of a KPZ surface depends {\em locally} on surface fluctuations, and hence cannot describe nonequilibrium surface dynamics with nonlocal interactions.} 

{ Theoretical studies on nonlocal interactions has a longstanding
history in 
equilibrium systems~\cite{leq1,leq2,leq3,alp,tb1,tb2}. 
  Examples of their prominent nonequilibrium counterparts include interface dynamics involving nonlocal interactions, e.g., flame front propagation, thin film growth~\cite{pelce}, and shading phenomena in surface growth~\cite{ketterl}.    Kinetic roughening in the presence of nonlocal interactions~\cite{somen} display generic non-KPZ scaling behavior.  Nonlocal effects are often important in biological growth processes; see, e.g., Ref.~\cite{nonlocal-eden} for a recent  study. Furthermore, in many  applications,  the growth is controlled by fast nonlocal transport not included in the KPZ equation. Prominent examples include diffusion-controlled nonlocal transport~\cite{nonloc1}, dissolution
or precipitation processes~\cite{nonloc2}, gas-solid reactions~\cite{nonloc3}, a
variety of reaction engineering processes~\cite{nonloc4},  diffusion-limited erosion, that displays nonlocal stabilization of surfaces~\cite{krug-prl} (see also Ref.~\cite{nicoli}), and even geological contexts, e.g., earth surface roughness~\cite{geo}. Inspired by these past studies, we explore the generic consequences of competition between local contributions and those that depend on the global surface profile, i.e., nonlocal contributions to the local surface velocity, by constructing a purpose-built conceptual model.}

In this Letter, we set up and study  a generalization of the KPZ equation, where the surface velocity depends, in contrast to the KPZ equation, {\em nonlocally} on the surface fluctuations. We do this by adding  symmetry-permitted nonlocal nonlinear gradient terms that are of the same order as the usual KPZ nonlinear term. These nonlocal, nonlinear terms have the same scaling as the usual local nonlinear term of the KPZ equation. This allows us to study competition and interplay between local and nonlocal nonlinear effects, resulting into stable steady states and roughening transitions distinct from both the usual KPZ equation, or the KPZ equation with truly long-range effects (with either long range nonlinearity or long range noises)~\cite{medina,somen}.
%
To generalize the scope of our study, we also include chiral contributions, 
which is ubiquitous in soft matter and biologically inspired systems; see, e.g., Refs.~\cite{chiral-mem1,chiral-mem2}.
The resulting equation in 2D, named active-KPZ or a-KPZ equation, is 
\begin{eqnarray}
 \frac{\partial h}{\partial t} = &\nu \nabla^2 h+\frac{\lambda}{2} (\boldsymbol\nabla h)^2 + \lambda_1 Q_{ij}({\bf r})(\nabla_ih \nabla_jh) \nonumber \\ 
 & + \lambda_2 Q_{ij}({\bf r})e_{jm}(\nabla_i h \nabla_m h)+\eta,  \label{ch-kpz}  
\end{eqnarray}
a {\em nonlocal} generalization to the usual KPZ equation that is distinct from the one considered in Ref.~\cite{somen}. Here, the tensor $e_{jm}$ is the 2D totally antisymmetric matrix. Further, $Q_{ij}({\bf r})$ is the longitudinal projection operator that in the Fourier space is $Q_{ij}({\bf k})=k_ik_j/k^2$, where ${\bf k}$ is a Fourier wave vector, and is nonlocal. Physically, $\lambda_1 Q_{ij}({\bf r})(\nabla_ih \nabla_jh)+\lambda_2 Q_{ij}({\bf r})e_{jm}(\nabla_i h \nabla_m h)$ is the contribution to the surface velocity { normal to the base plane} $v_p=\partial h/\partial t$ that is {\em nonlocal} in height fluctuations ${\boldsymbol\nabla} h$. 
Noise $\eta$ is a zero-mean, Gaussian-distributed white noise with a variance
 $\langle \eta({\bf x},t)\eta({\bf 0},0) \rangle = 2D \delta^d({\bf x}) \delta(t)$. 
 We extract the scaling of the stable phases, which exists for a range of the model parameters. In particular, we show that the variance $\Delta\equiv \langle h({\bf x},t)^2\rangle\sim [\ln(L/a)]^\mu$ for a surface of lateral size $L$, where $\mu<(>)1$ for sub (super) logarithmic roughness and $a$ is a microscopic cutoff. This defines positional generalized  quasi-long-ranged order (QLRO), generalizing the well-known QLRO of EW surfaces~\cite{stanley},  in which $\Delta\sim \ln (L/a)$, i.e., $\mu=1$.  Further, the time-scale of relaxation $\tau(L)\sim L^2\,(\ln\,L)^{-\kappa},\,\kappa>0$, i.e., logarithmically  superdiffusive. Both $\mu$ and $\kappa$ are nonuniversal. They vary continuously with $\lambda_1/\lambda,\,\lambda_2/\lambda$.

 The form of Eq.~(\ref{ch-kpz}) can be obtained by first considering the  mapping from the KPZ equation to the Burgers equation~\cite{forster} in terms of the ``Burgers velocity ${\bf v}={\boldsymbol\nabla} h$.'' Now generalizing the Burgers equation nonlinearity $\lambda{\boldsymbol\nabla} v^2$ to $\lambda_1\nabla_j (v_i v_j)+\lambda_2 e_{jm}\nabla_j (v_i v_m)$, and then writing them in terms of $h$ produces the $\lambda_1$ and $\lambda_2$ terms in (\ref{ch-kpz}); see the Supplemental Material (SM)~\cite{supple}. 

The $\lambda_1$ and $\lambda_2$ terms in (\ref{ch-kpz}) can be motivated by considering a nearly flat nonequilibrium surface without any momentum conservation described by a single valued height field $h({\bf x},t)$ in Monge gauge ~\cite{chaikin,nelson-book}, with an active conserved density $\rho({\bf x},t)$ living on it. Its hydrodynamic equation, retaining only the lowest order in nonlinearities and spatial gradients, reads
\begin{eqnarray}
 \frac{\partial h}{\partial t}= \nu\nabla^2 h +\frac{\lambda}{2}({\boldsymbol\nabla}h)^2 + v(\rho) + \eta,\label{gen-eq}
 \end{eqnarray}
 where $v(\rho)$ is a local density-dependent velocity of the membrane; $v(\rho)=v_0 +g_1\rho$ to the leading order in $\rho$; $g_1$ is a coupling constant of either sign.
 Further, density $\rho$ follows $\partial_t\rho = - {\boldsymbol\nabla}\cdot {\bf J}$, where $\bf J$ is the current. The specific form of the particle dynamics decides the structure of $\bf J$. We choose ${ J_i}=-\overline D{\nabla_i}\rho + {\nabla_j}\sigma_{ij}$,  where $\sigma_{ij}\equiv \alpha\nabla_i h\nabla_j h + \beta e_{jm} \nabla_i h \nabla_m h$ is reminiscent of ``active stresses'' found in active matter theories~\cite{sriram-RMP}, the $\beta$ term is a chiral contribution. The quadratic dependence of $\bf J$ on ${\boldsymbol\nabla} h$ implies the active particles (i) respond, unsurprisingly, to the height fluctuations, but not the  absolute height; and (ii) ignoring gravity, the particles do not distinguish valleys from the hills (although the surface itself breaks the inversion symmetry). Here, $\overline D>0$ is a diffusivity. We focus on the quasistatic limit of infinitely fast dynamics of $\rho$, such that $\partial\rho/\partial t\approx 0$, giving $\overline D\nabla^2 \rho =\alpha \nabla_i\nabla_j (\nabla_i h\nabla_jh)+\beta e_{jm}\nabla_i\nabla_j(\nabla_i h\nabla_m h)$ neglecting any noise in the $\rho$ dynamics. Now use this to eliminate $\rho$ in (\ref{gen-eq}) to get  (\ref{ch-kpz}), after absorbing a factor of $\overline D$. (We have implicitly assumed $\alpha,\beta$ to scale with $\overline D$, and ignored any advective-type nonlinearity originating from projecting the particle dynamics on the plane of the membrane in the large $\overline D$ limit). 
 All of $\lambda,\lambda_1,\lambda_2$ can be individually positive and negative. The chiral term is 2D specific; the other two nonlinear terms with coefficients $\lambda$ and $\lambda_1$  can exist in any dimension $d$.  Thus the $\lambda_1$ and $\lambda_2$ terms in (\ref{ch-kpz}) are physical, although our active species origin need not be the only possible source of these two terms.  See Ref.~\cite{frey-astik} for a similar mechanism to generate an effective nonlocal dynamics in the noisy Fisher-Kolmogorov equation~\cite{fisher1,fisher2} for population dynamics coupled with a fast chemical signal. { Further Eq.~(\ref{ch-kpz}) can be realized microscopically by considering an ``active'' 2D single-step model for a 2D KPZ surface with point particles living on it. The dynamical update rules of the modified single-step model now depend on the local excess or deficit population of the active particles, instead of being constants as they are in standard single-step models~\cite{sng1,sng2,sng3,sng4,sng5}. The particle hopping rates to the nearest neighbor sites in turn depends not only on the number inhomogeneities, but also on the height fluctuations (but without distinguishing local valleys from hills). Monte-Carlo simulations of this model, focusing on the limiting case of fast dynamics by the number fluctuations, should bring out the physics described in this Letter. The limit of fast particle dynamics can be implemented by considering time-scale separations in the rates of particle position updates and surface conformation updates.} 

At one dimension, the $\lambda_2$ term vanishes, and the $\lambda_1$ term becomes indistinguishable from the $\lambda$-term. 
The transformation $x_i'=x_i-(\lambda+2\lambda_1)c_it-\lambda_2e_{ij}c_{j}t$ and $t'=t$, together with the height function $h$ transforming as $h'({\bf x}',t')=h({\bf x},t)+\textbf{c}\cdot\textbf{x},$ leaves Eq.~(\ref{ch-kpz})  invariant; see the SM~\cite{supple}. This generalizes invariance of the usual KPZ equation under a pseudo-Galilean transformation~\cite{stanley}. 


{ Similar to the KPZ equation, dimensional analysis via scaling ${\bf r}\rightarrow b{\bf r},\,t\rightarrow b^zt,\, h\rightarrow b^\chi h$,  where $z$ and $\chi$ are the dynamic and roughness exponents, 
reveals that 
all of $\lambda,\lambda_1,\lambda_2$ scale similarly, and hence are equally relevant (in the scaling sense). Furthermore,
$d=2$ is the critical dimension of Eq.~(\ref{ch-kpz}); see the SM~\cite{supple}}. Whether it is the upper or lower critical dimension requires further analysis that follows below. 
That all of $\lambda,\lambda_1,\lambda_2$ scale the same way is important: it means the nonlocal, nonlinear effects in (\ref{ch-kpz}) are as relevant as the short-range, local nonlinear effects in the original KPZ equation~\cite{stanley}. This feature clearly distinguishes the active KPZ equation~(\ref{ch-kpz}) from generalized KPZ equations with genuine long-range interactions~\cite{somen}. Indeed, just as the usual KPZ equation~\cite{stanley} is universal in the sense that all short-range growth processes with just one soft mode (height $h$) and without any conservation laws, inversion symmetry and disorder should be described by it; the active KPZ equation~(\ref{ch-kpz}) should likewise describe all such nonlocal growth processes having the same scaling properties as the corresponding local growth processes in the KPZ equation, and with just one soft mode ($h$) but without any conservation laws, inversion symmetry, and disorder, highlighting the universal nature of (\ref{ch-kpz}).

We first determine if Eq.~(\ref{ch-kpz}) has a stable nonequilibrium steady state (NESS), and second, if so, the scaling properties in those NESS. We use renormalization group (RG) framework, well suited to systematically handle the diverging corrections encountered in na\"ive perturbation theories.  The Wilson dynamic RG method  for our model closely resembles that for the KPZ equation~\cite{kpz,stanley,forster,tauber}; see the SM~\cite{supple} for the one-loop Feynman diagrams. 
%
There are no one-loop corrections to $\lambda,\lambda_1,\lambda_2$. However, there are diverging one-loop corrections to $\nu$ and $D$. Dimensional analysis allows us to identify an effective dimensionless coupling constant $g$ and two dimensionless ratios $\gamma_1,\,\gamma_2$ defined as
 $g\equiv \frac{\lambda^2 D}{\nu^3}\frac{1}{2\pi},\,\gamma_1\equiv \frac{\lambda_1}{\lambda},\,\gamma_2\equiv \frac{\lambda_2}{\lambda}$. 
The  RG recursion relations for $D,\,\nu$ at the one-loop order (here $l$ is the ``RG time;'' $\exp(l)$ is a length scale)
\begin{align}
   &\frac{dD}{dl}=D\Bigr[z-d-2\chi+g{\cal B}(\gamma_1,\gamma_2)\Bigr], \label{D-flow} \\
   &\frac{d\nu}{dl}=\nu\Bigr[z-2+g{\cal C}(\gamma_1,\gamma_2) \Bigr], \label{nu-flow}
\end{align}
with $\gamma_1,\,\gamma_2$ being marginal at the one-loop order, stemming from the nonrenormalization of $\lambda,\lambda_1,\lambda_2$ at that order. 
Here, ${\cal B}[\gamma_1, \gamma_2]=\frac{3}{8}\gamma_1^2+\frac{1}{2}\gamma_1+\frac{1}{8}\gamma_2^2+\frac{1}{4},\,{\cal C}[\gamma_1, \gamma_2]=\frac{1}{2}\gamma_1^2+\frac{5}{8}\gamma_1+\frac{1}{8}\gamma_2^2$; ${\cal B}>0$. Flow equations (\ref{D-flow}) and (\ref{nu-flow})  yield the flow equation for $g$:
\begin{equation}
 \frac{dg}{dl}=-g^2{\cal A}[\gamma_1,\gamma_2],\label{g-flow}
\end{equation}
where 
 ${\cal A}[\gamma_1,\gamma_2]=
 \frac{9}{8}\gamma_1^2+\frac{11}{8}\gamma_1+\frac{1}{4}\gamma_2^2-\frac{1}{4}$. 

\begin{figure}[b!]
\centering \includegraphics[width=0.48\textwidth]{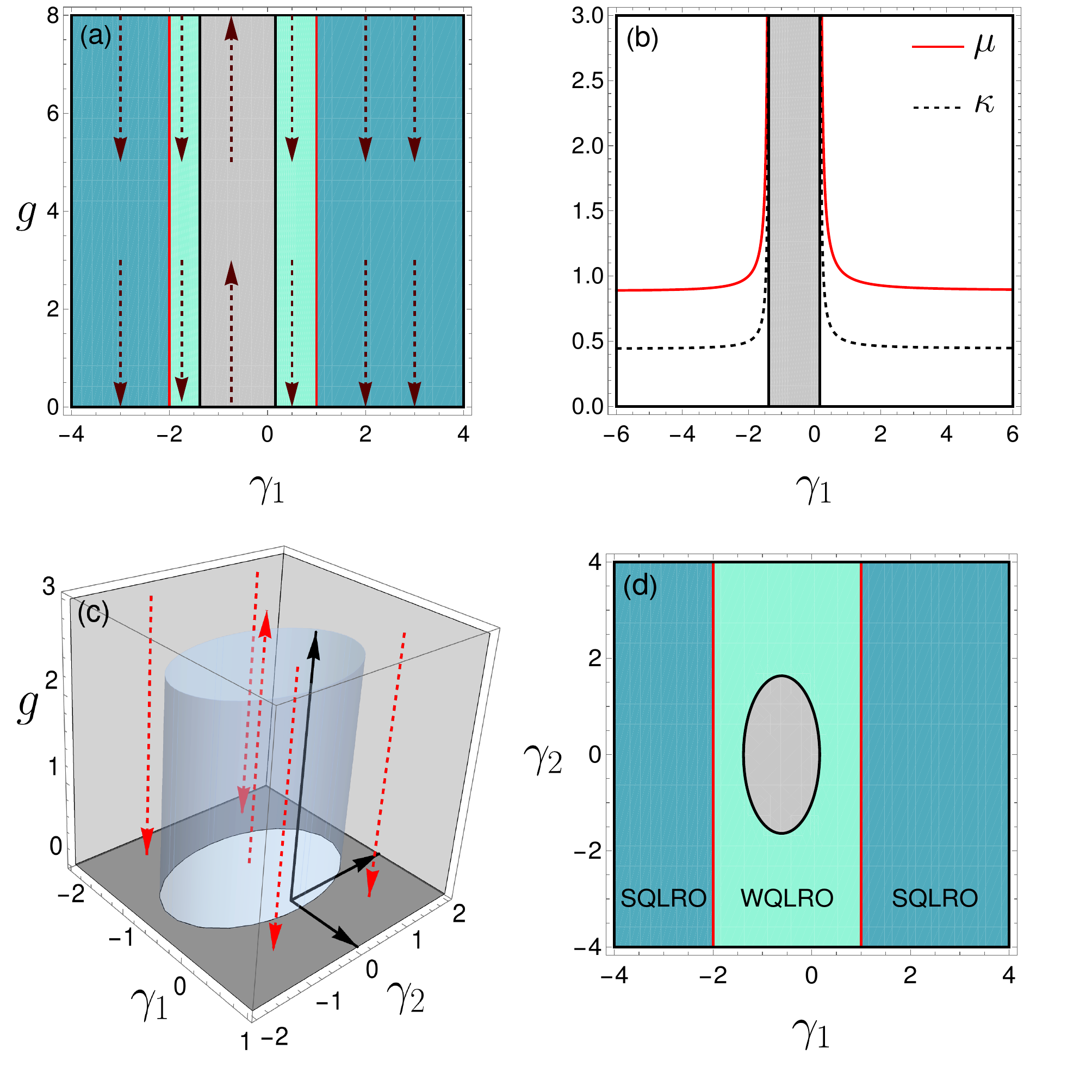}
\caption{(a) RG flow diagram in the $g$-$\gamma_1$ plane in the achiral limit ($\gamma_2=0$). Arrows indicate RG flows. Flow in the stable (unstable), i.e., toward (away from), $g=0$ {region} are marked.  (b) Variations of $\mu$ and $\kappa$ as functions of $\gamma_1$ in the stable region for the achiral case. (c) { RG flow diagram in the space spanned by $\gamma_1$-$\gamma_2$-$g$ in the full a-KPZ equation.} 
RG flow lines in the stable and unstable regions are shown by the arrows. (d) Phase diagram { in the $\gamma_1$-$\gamma_2$ plane for the a-KPZ equation}. The central gray region containing the origin is unstable.  Regions with SQLRO and WQLRO are marked (see text).}
\label{flow}
\end{figure}

In the achiral case, i.e., $\gamma_2=0$, an RG flow diagram in the $g$-$\gamma_1$ plane is shown in Fig.~\ref{flow}(a). The condition ${\cal \tilde A}(\gamma_1)\equiv {\cal A}(\gamma_1,\gamma_2=0)=0$ defines two solid (black) lines $\gamma_1=\gamma_+,\,\gamma_-$ parallel to the $g$ axis in the $g$-$\gamma_1$ plane, where $\gamma_+=0.161,\gamma_-=-1.383$, such that for $\gamma_+>\gamma_1>\gamma_-$ { (gray region)}, the RG flow lines flow away parallel to the $g$ axis toward infinity, indicating a perturbatively inaccessible, presumably rough, phase with short-ranged positional order. In this unstable region $g(l)$ diverges as $l\rightarrow 1/[|{\cal \tilde A}(\gamma_1)|]$,  reminiscent of the 2D KPZ equation~\cite{stanley}, presumably corresponding to algebraically rough phase~\cite{bmhd1,2d-kpz-expo}. Outside this region, where ${\cal \tilde A}(\gamma_1)>0$, the flow lines flow toward $g=0$ parallel to the $g$ axis, implying stability, while $g(l)\approx 1/[l{\cal \tilde A}(\gamma_1)]$ vanishes slowly in the long wavelength limit $l\rightarrow \infty$. Although $g^*=0$ is the only fixed point (FP) in the stable region, the vanishing of $g(l)$ is so slow, being proportional to $1/l$, the parameters $D$ and $\nu$ are infinitely renormalized, altering the linear theory scaling in the long wavelength limit. The simplest way to see this is to set $z=2,\,\chi=0$ (i.e., their linear theory values) in  (\ref{D-flow}) and (\ref{nu-flow}) with $\gamma_2=0$, which gives 
\begin{equation}
 D(l)=D_0 l^{{\cal \tilde B}/{\cal\tilde A}},\, \nu(l)=\nu_0l^{{\cal \tilde C}/{\cal\tilde A}},
\end{equation}
where ${\cal\tilde B}(\gamma_1)\equiv {\cal B}(\gamma_1,\gamma_2=0),\,{\cal\tilde C}(\gamma_1)\equiv {\cal C}(\gamma_1,\gamma_2=0)$, $D_0,\nu_0$ are the small-scale or unrenormalized values of $D$ and $\nu$.
Since ${\cal \tilde B}$ is positive definite, $D(l)\gg D_0$ for $l\rightarrow \infty$. { On the other hand, ${\cal \tilde C}$ is positive in stable regions, for which  $\nu(l)\gg \nu_0$ for $l\rightarrow \infty$}, giving the time-scale $\tau(L)\sim L^2[\ln (L/a)]^{-\kappa}$ for relaxation over lateral size $L$, where $\kappa (\gamma_1)= {\cal \tilde C}/{\cal \tilde A}$ is a positive definite but nonuniversal, $\gamma_1$-dependent exponent. 
The logarithmic modulation in $\tau(L)$ implies (i) breakdown of conventional dynamic scaling~\cite{activexy1,activexy2,john2dinterface}, and (ii) {\em nonuniversally faster} relaxation, being parametrized by $\gamma_1$, of fluctuations. Furthermore, by defining RG time $l\simeq\ln(1/{aq})$  and using  $\nu(q), D(q)$, the variance is
\begin{align}
\Delta\equiv\langle h^2({\bf x},t)\rangle \sim \int_{1/L}^{1/a}d^2q\frac{D(q)}{\nu(q) q^2}\sim [\ln (L/a)]^{\mu},\label{varih}
\end{align}
where $\mu (\gamma_1)=1+ ({\cal \tilde B}-{\cal\tilde C})/{\cal \tilde A}$ is also nonuniversal, parametrized by $\gamma_1$, and can be more or less than unity, depending upon the sign of ${\cal \tilde B}-{\cal\tilde C}$, as mentioned above. Variations of $\mu$ and $\kappa$ as functions of $\gamma_1$ are shown in Fig.~\ref{flow}(b). For $\mu(\gamma_1)<1(>1)$, $\Delta(\gamma_1)$ grows with the system size $L$ slower (faster) than positional QLRO, as in the 2D EW equation~\cite{ew}. We call these stronger (weaker) than QLRO or SQLRO (WQLRO), corresponding to sub (super) logarithmically rough surfaces with positional generalized QLRO, that generalize the well-known QLRO in the 2D EW equation or 2D equilibrium XY model~\cite{chaikin}. In particular, the minimum of $\mu=0.89$. In Fig.~\ref{flow}(a) the blue outer regions (green inner strips) correspond to SQLRO (WQLRO).  Solid red lines correspond to positional QLRO. These 
results are reminiscent of the logarithmic anomalous elasticity in three-dimensional equilibrium smectics~\cite{pelc1,pelc2}, and  a 2D equilibrium elastic sheet having vanishing thermal expansion coupled with Ising spins~\cite{sm1,sm2}; see also Ref.~\cite{john2dinterface} for similar results.

Including the chiral effects ($\gamma_2\neq 0$), 
stability of the RG flow is now determined by ${\cal A}(\gamma_1,\gamma_2)>0$. Flow lines having initial conditions within a narrow elliptical cylinder, containing the origin (0,0,0), and having the axis parallel to the $g$ axis, with its surface given by ${\cal A}(\gamma_1,\gamma_2)=0$ for any $g$, run away parallel to the $g$ axis, leaving the perturbatively accessible region. Flow lines with initial conditions falling in regions outside of this elliptical cylinder flow toward the $\gamma_1$-$\gamma_2$ plane with stable states.  See Fig.~\ref{flow}(c) depicting the RG flow lines in the space spanned by $\gamma_1$-$\gamma_2$-$g$. Outside the elliptical cylinder $g(l)\sim 1/({\cal  A}l)$ for large $l$, similar to its achiral analog. Inside the cylinder, $g(l)$ diverges as $l\rightarrow 1/(|{\cal A}|)$ from below. Focusing on the $\gamma_1$-$\gamma_2$ plane,  ${\cal A}(\gamma_1,\gamma_2)=0$
 sketches out an inner elliptical unstable region, whereas the outer region is stable; see Fig.\ref{flow}(d). 
We use the above results to find that in the stable region $\Delta\sim [\ln (L/a)]^{\mu(\gamma_1,\gamma_2)}$, where $\mu=1+ ({\cal B}-{\cal C})/{\cal A}$ is now parametrized by both $\gamma_1,\gamma_2$. Similar to and quantitatively extending the achiral case,  $\mu<1 (>1)$ is referred to as SQLRO (WQLRO), giving positional generalised QLRO. The SQLRO and WQLRO regions are demarcated within the stable region in Fig.~\ref{flow}(d). 



The equal-time height-difference correlator 
$C_h(r,0)\equiv\langle [h({\bf x},t)-h({{\bf 0}},t)]^2\rangle\sim\frac{D_0}{\nu_0}[\ln (r/a)]^\mu$, for large $r\equiv |{\bf x}|\gg a$,
indicating logarithmically faster or slower rise with the separation $r$ for large $r$~\cite{activexy1,activexy2}, again generalizing 
the well-known QLRO found in a 2D EW surface.


Our continuously varying scaling exponents are a crucial outcome of the nonrenormalization of $\lambda,\lambda_1$ and $\lambda_2$, rendering $\gamma_1,\gamma_2$ marginal, which have been demonstrated at the one-loop order. Unlike the usual KPZ equation, Galilean invariance of the present model ensures nonrenormalization of a combination of $\lambda,\,\lambda_1,\,\lambda_2$, and not each of them individually. Thus,  there is no surety that 
$\gamma_1,\gamma_2$ should remain marginal even at higher-loop orders. We now argue that these possible higher-loop contributions, even though they may exist, actually do not matter. For large $l$, $g(l)\sim 1/l$  at the one-loop order. At higher-loop orders, the Feynman diagrams will contain higher power of $g$. Hence, a general scaling solution for $g(l)$ should have the form $g(l)\sim 1/l +\sum_n c_n/l^n,$ and $n>1$ is an integer.  Thus, the higher-loop corrections to the one-loop
solution of $g(l)$ should vanish like $1/l^s,\,s>1$. Therefore, their integrals over $l$ from zero to infinity will be finite, so they will not change the anomalous
behavior of $D$ and $\nu$.  Similarly, they cannot make any
divergent contribution to $\gamma_1(l)$ and $\gamma_2(l)$, even though
there can be higher-loop diagrams. Therefore, our one-loop results are, in fact, asymptotically exact. This then implies that the continuous variation of the scaling
exponents, making them nonuniversal, is also asymptotically exact in the long wavelength limit. See Refs.~\cite{activexy1,activexy2,niladri1,bmhd1,bmhd2,frey-pnas,chate,astik-qckpz} for similar nonuniversal scaling exponents in other models.


At higher dimensions $d>2$, the chiral term with coupling $\lambda_2$ cannot exist. 
The other two achiral  nonlinear terms in  Eq.~(\ref{ch-kpz}) are present at $d>2$.  The RG recursion relations for $d>2$ can be obtained from the Feynman diagrams given in the SM~\cite{supple} with $\gamma_2=0$. Using a $d=2+\epsilon$ expansion as in the KPZ equation~\cite{natter}, we find at the one-loop order or  to the lowest order in $\epsilon,$
\begin{equation}
 \frac{dg}{dl}=-\epsilon g - {\cal \tilde A}(\gamma_1) g^2.\label{higher-g-flow}
\end{equation}
Parameter $\gamma_1$ remains marginal at the lowest order.
Therefore, if ${\cal \tilde A}(\gamma_1)>0$, $g(l)$ flows to zero rapidly, with $g(l)\sim g(0)\exp (-\epsilon\, l)$ in the long wavelength limit; $g^*=0$ is the only FP that is globally stable. This renders the nonlinearities irrelevant in the RG sense. Therefore, scaling in the long wavelength limit is identical to that in the EW equation: $z=2,\,\chi=(2-d)/2$. Furthermore, $d=2$ is then the  upper critical dimension. On the other hand, if  ${\cal \tilde A}(\gamma_1)<0$, $g(l)$ has three FPs: $g^*_c=-\epsilon/ {\cal \tilde A}(\gamma_1)$, an unstable FP, parametrized by $\gamma_1$ and separating possibly two stable FPs,  { one being at $g^*=0$ Gaussian FP with EW scaling}, and another putative perturbatively inaccessible FP, corresponding presumably to an algebraically rough phase. This gives, with 2D as the lower critical dimension, a roughening transition at $d>2$, very similar to the KPZ equation at $d>2$, but with one caveat. 
At this unstable FP, using (\ref{D-flow}) and (\ref{nu-flow}), to ${\cal O}(\epsilon)\,$
${z}=2+\epsilon\frac{{\cal \tilde C}(\gamma_1)}{{\cal \tilde A}(\gamma_1)},\,\, \chi=-\epsilon \frac{{\cal \tilde C}(\gamma_1)}{{\cal \tilde A}(\gamma_1)}$,
depend explicitly on $\gamma_1$ and deviate from their linear theory (or EW equation) values already at ${\cal O}(\epsilon)$.
This is in contrast to the KPZ equation at $d>2$, where $z$ and $\chi$ at the unstable FP are at least ${\cal O}(\epsilon^2)$~\cite{natter}.   In fact, application of the Cole-Hopf transformation shows that $z=2,\,\chi=0$ at the unstable FP of the KPZ equation at $d>2$~\cite{cole}.

\begin{figure}[h]
\centering \includegraphics[width=0.48\textwidth]{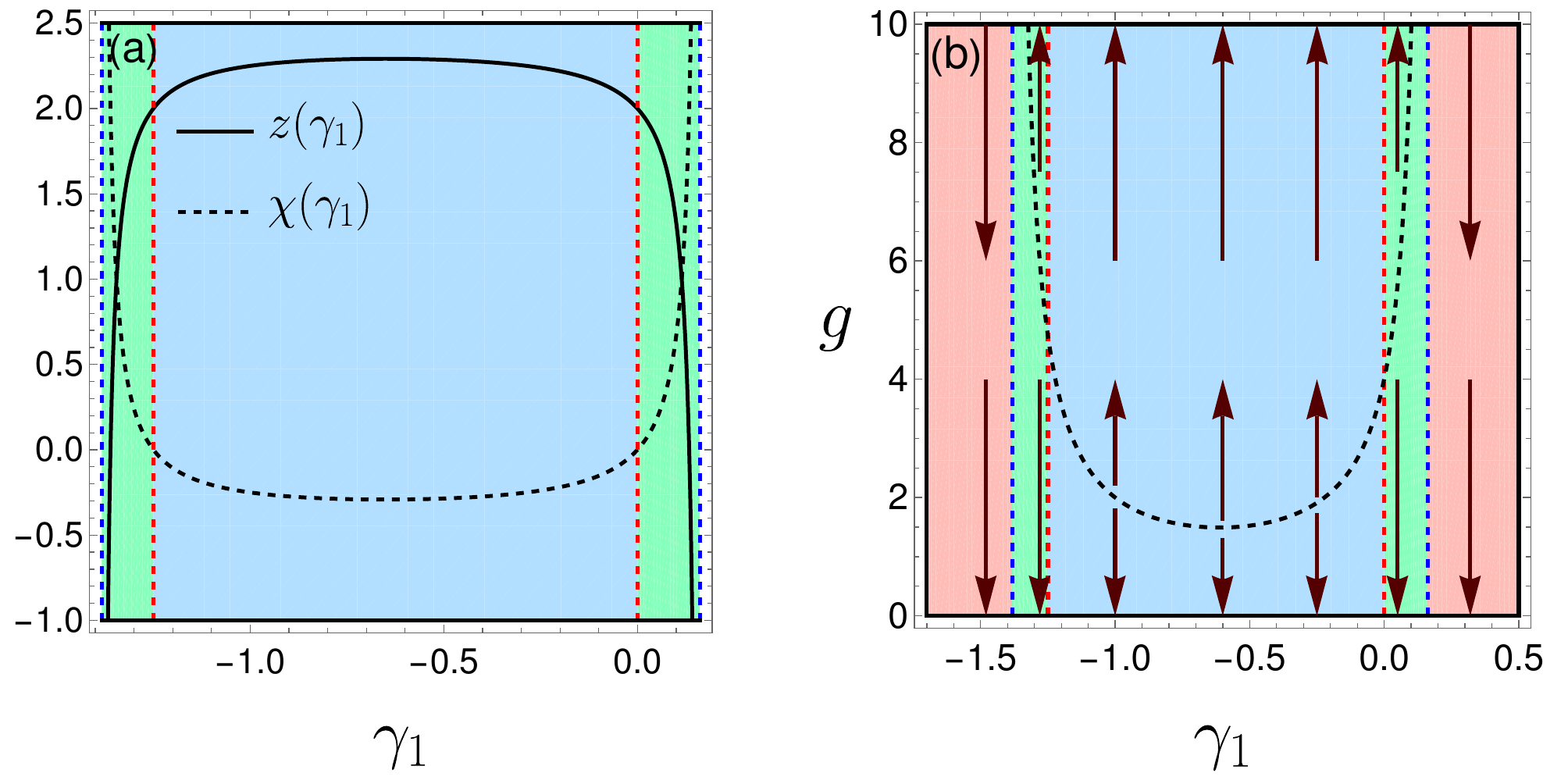}
\caption{(a) Variation of $z$ and $\chi$ with $\gamma_1$ on the fixed line $ g_c^*=-1/ {\cal \tilde A}(\gamma_1)$ for $\epsilon=1.$ (b) RG flow diagram in the $g$-$\gamma_1$ plane for $d>2$. The black dashed line is the fixed line $ g_c^*=-1/ {\cal \tilde A}(\gamma_1)$ for $\epsilon=1$, bounded by lines $\gamma_1=\gamma_-,\gamma_+$ (blue dashed lines). Stable (unstable) flow lines are the arrows pointing toward (away from) $g=0$ (see text).}
%
\label{gamm-g-flow}
\end{figure}

 When ${\cal \tilde A}(\gamma_1)<0$, the solution of ${\cal \tilde C}(\gamma_1)=0$ gives the red dashed lines $\gamma_1=\gamma^-,\gamma^{\dag}$ where, $\gamma^{\dag}=0$ and $\gamma^- =-1.25$; see Fig.~\ref{gamm-g-flow}(a) for a variation of $z$ and $\chi$ with $\gamma_1$ for a fixed $\epsilon$.  Green strips  correspond to  ${\cal \tilde C}(\gamma_1)>0$ where $\chi>0$  and $z<2$;  ${\cal \tilde C}(\gamma_1)<0$ is the blue region where $\chi<0$ and $z>2$.  For  a given  $\epsilon$, maximum value of $z$ and minimum value of $\chi$ are $z_\text{max}=2+0.292\epsilon$ and $\chi_\text{min}=-0.292\epsilon$ at $\gamma_1=-0.651$, such that the dynamics is slowest and the surface is smoothest at the unstable FP. 
{ Since $\chi_\text{min}>\chi_\text{EW}=-\epsilon/2$, an a-KPZ surface at the unstable FP is always rougher than an EW surface.}

{
In the $g$-$\gamma_1$ plane, $ g_c^*=-\epsilon/ {\cal \tilde A}(\gamma_1)$  is a fixed line,  such that RG flow lines with initial $g$ values above the  line flows to perturbatively inaccessible FP, see Fig.~\ref{gamm-g-flow}(b).} And for systems with  initial $g$ values lying below the    line, the RG flow lines run parallel to the $g$ axis toward  { Gaussian FP}, corresponding to the smooth phase belonging to the EW class. This behavior holds within a range $\gamma_->\gamma_1>\gamma_+$. As $\gamma_1\rightarrow \gamma_+,\,\gamma_-$, ${\cal \tilde A}(\gamma_1)$ vanishes and $g_c^*$ diverges. As soon as $\gamma_1$ exceeds $\gamma_+$ or falls short of $\gamma_-$, $g_c^*$ no longer exists with the roughening transition disappearing. RG flow lines starting from any initial condition with $\gamma_1>\gamma_+$ or $\gamma_1<\gamma_-$ (red region) where ${\cal \tilde A}(\gamma_1)>0$, flow  to $g^*=0$ ensuing scaling belonging to the EW class. 



 In summary, we have proposed and studied an ``active KPZ'' equation, having a surface velocity $v_p$ depending nonlocally on the surface gradients. Surprisingly,  we find { stable surfaces with positional generalized QLRO or generalized logarithmic roughness} with nonuniversal exponents for  wide-ranging choices of the model parameters, unlike the 2D KPZ equation. Physically, this is due to the competition between the nonlocal and local nonlinear terms and the lack of their renormalization. Indeed, this competition between the nonlocal and local nonlinear terms distinguishes our model~(\ref{ch-kpz}) from that studied in Ref.~\cite{somen}, giving either generalized QLRO with nonuniversal scaling exponents or a novel roughening transition even in 2D controlled by the relative strengths of the local and nonlocal interactions. At $d>2$, sufficiently strong nonlinear nonlocal effects can either entirely suppress the KPZ roughening transitions, resulting into only smooth surfaces, or else give a roughening transition with nonuniversal scaling very different from the well-known roughening transition in the KPZ equation. 
 Heuristically, a nonlocal part in $v_p$ means a local large fluctuation can generate a propulsion not just locally, but over large scales, which when sufficiently strong can suppress local variations in $v_p$ due to the local KPZ-nonlinear term. This in turn has the effect of reducing surface fluctuations. For other parameter choices, a KPZ-like  perturbatively inaccessible rough phase is speculated. This may be explored by mode-coupling methods~\cite{jkb-mct}. In that parameter space, the roughening transition survives at $d>2$, but with significantly different scaling properties, { again with nonuniversal exponents}. 
 We hope our studies here will provide further impetus to study nonlocal effects on similar nonequilibrium surface dynamics models, e.g.,  the conserved KPZ~\cite{ckpz,ckpz+} and the $|{\bf q}|$KPZ~\cite{qkpz-paper} equations. 

A.B. thanks  the SERB, DST (India) for partial financial support through the MATRICS scheme [File No. MTR/2020/000406].

\bibliography{qkpzddim.bib}

\begin{thebibliography}{62}
\bibitem [{sup()}]{supple}%
  \BibitemOpen
  \bibfield  {title} {\enquote {\bibinfo {title} {\uppercase{S}ee
  \uppercase{S}upplemental \uppercase{M}aterial for the derivation of
  \uppercase{E}q.~(\ref{ch-kpz}); the \uppercase{G}alilean invariance of
  \uppercase{E}q.~(\ref{ch-kpz}); its critical dimension; and the one loop
  \uppercase{F}eynman diagrams for the renormalization group calculations.
  \uppercase{I}t also contains
  \uppercase{R}efs.~\cite{stanley,forster,chaikin,nelson-book,tauber,activexy1,activexy2,sm1,helm1,helm2,bausch,yakhot}.}}\
  }\href@noop {} {\ }\BibitemShut {NoStop}%
\end{thebibliography}

\clearpage
\onecolumngrid
\begin{center}
\textbf{\LARGE Supplemental Material}
\end{center}
\vspace{0.5cm}

\section{Active KPZ equation} 

We first construct the active KPZ equation in terms of the vector field ${\bf v}({\bf r},t)\equiv {\boldsymbol\nabla}h$. Field $\bf v$ has a definite physical interpretation:  for an active membrane $\bf v$ is the  (small) fluctuation of the local normal to the membrane surface measured in the Monge gauge~\cite{nelson-book,chaikin}. Noting that the Burgers velocity $\bf v$ is a conserved vector field, it must follow a generic conservation law of the form
\begin{equation}
 \frac{\partial v_i}{\partial t}=-\nabla_jJ_{ij},
\end{equation}
where $J_{ij}$ is the velocity current. 
In general $J_{ij}$ can be decomposed as the sum of a symmetric part $J_{ij}^s=J_{ji}^s$ and an antisymmetric part $J_{ij}^a=-J_{ji}^a$: $J_{ij}=J_{ij}^s+J_{ij}^a$. Since $\bf v$ is purely irrotational, it can be expressed, via the Helmholtz theorem~\cite{helm1,helm2}, solely in terms of its divergence, i.e., ${\boldsymbol\nabla}\cdot {\bf v}\equiv {\cal D}$, which follows the equation
\begin{equation}
 \frac{\partial {\cal D}}{\partial t}= - \nabla_i\nabla_j J_{ij} = -\nabla_i\nabla_j J_{ij}^s.
\end{equation}
Thus, $J_{ij}^a$ plays no role in the dynamics of $\cal D$ and hence of $\bf v$. Without any loss of generality, we then set $J_{ij}^a=0$. We further express $J_{ij}=J_{ij}^s$ as
\begin{equation}
 J_{ij}=\frac{1}{2}(\partial_i\mu_j+\partial_j\mu_i),
\end{equation}
where ${\boldsymbol\mu}$ is the ``chemical potential'' vector, which in general can have irrotational and solenoidal parts. The latter part does not contribute to the dynamics of ${\cal D}$ and hence of $\bf v$, and hence can be set to zero, leaving $\boldsymbol\mu$ purely irrotational. In that case,
\begin{equation}
 \frac{\partial {\bf v}}{\partial t}=-\nabla^2 {\boldsymbol\mu}.
\end{equation}
The Burgers equation is given by
\begin{equation}
 \frac{\partial  v_i}{\partial t}=\nu\nabla^2 v_i +\frac{\lambda}{2}\nabla_i v^2 + f_i.
\end{equation}
Therefore, in the linearised Burgers equation  ($\lambda=0$), ${\boldsymbol\mu}=-\nu {\bf v}$, a local quantity, whereas for the Burgers equation
${\boldsymbol\mu}=-\frac{\lambda}{2}\nabla^{-2}{\boldsymbol\nabla} {v^2}$, a nonlocal quantity.

Now consistent with the conservation law form of the Burgers equation, $\mu_i$ in the Burgers equation can be generalised. It also admits, at the same order, a second contribution of the form $\sim \nabla_j (v_i v_j)/\nabla^2$. Indeed, including this term the most general equation of $\bf v$ retaining only up to the lowest order in nonlinearities and spatial gradients that now includes a chiral contribution is of the form
\begin{equation}
 \frac{\partial {v_i}}{\partial t}=\nu\nabla^2 { v_i} + \frac{\lambda}{2}{\nabla_i}v^2+ \lambda_1 \nabla_j(v_i v_j)+ \lambda_2 \nabla_j e_{jm}(v_i v_m)+ f_i.\label{gen-burg}
\end{equation}
Both the additional nonlinear terms make nonlocal contributions to $\mu$.
Equation~(\ref{gen-burg}) reduces to the well-known Burgers equation~\cite{forster} when $\lambda_1=0=\lambda_2$. Further, the tensor $e_{ij}$ is the 2D totally antisymmetric matrix, with $e_{11}=0=e_{22}$ and $e_{12}=1=-e_{21}$. Thus, the $\lambda_2$-term in Eq.~(\ref{gen-burg}) is the chiral contribution. Now, further demanding that ${\bf v}$ is fully irrotational with ${\bf v}={\boldsymbol\nabla}h$,  the ``superfluid velocity'' of an XY model, or the deviation of the local normal of an interface in the Monge gauge~\cite{nelson-book}, we obtain,
\begin{eqnarray}
 \frac{\partial h}{\partial t} = &\nu \nabla^2 h+\frac{\lambda}{2} (\boldsymbol\nabla h)^2 + \lambda_1 Q_{ij}({\bf r})(\nabla_ih \nabla_jh) + \lambda_2 Q_{ij}({\bf r})e_{jm}(\nabla_i h \nabla_m h)+\eta,  \label{ch-kpz_supple}
\end{eqnarray}
where formally $Q_{ij}=\nabla_i\nabla_j/\nabla^2$ and is to be understood in terms of its Fourier transform.

\section{Galilean invariance}

Under the transformation $x_i'\rightarrow x_i-(\lambda+2\lambda_1)c_it-\lambda_2e_{ij}c_{j}t$ and $t'\rightarrow t$, where $i$ and $j$ can take values $1$ or $2$ we have,
\begin{align}
&\frac{\partial}{\partial t'}\rightarrow \frac{\partial}{\partial t}+(\lambda+2\lambda_1)\textbf{c}\cdot\boldsymbol\nabla-\lambda_2e_{ij}c_i\frac{\partial}{\partial x_j},\\
&\boldsymbol\nabla'\rightarrow \boldsymbol\nabla.
\end{align} 
Using these we show below that if the height function $h({\bf x},t)$ transforms as $h'({\bf x}',t')\rightarrow h({\bf x},t)+\textbf{c}\cdot\textbf{x}$, then Eq.~(\ref{ch-kpz}) remains  invariant. Considering RHS of Eq.~(\ref{ch-kpz}), the first term transforms as,
\begin{align}
\nu \nabla'^2 h'\rightarrow \nu \nabla^2 h.
\end{align} 
Second term transforms as,
\begin{align}
\frac{\lambda}{2} (\boldsymbol\nabla' h')^2\rightarrow \frac{\lambda}{2} (\boldsymbol\nabla h)^2+\lambda \boldsymbol \nabla h \cdot \textbf{c}+\frac{\lambda}{2}c^2.
\end{align} 
Third term transforms as,
\begin{align}
\lambda_1 \frac{\nabla'_i \nabla'_j}{\nabla'^2}(\nabla'_ih' \nabla'_jh')\rightarrow \lambda_1 \frac{\nabla_i \nabla_j}{\nabla^2}(\nabla_ih \nabla_jh)+2\lambda_1 \textbf{c}\cdot\boldsymbol\nabla h.
\end{align}
Chiral term transforms as,  
\begin{align}
\lambda_2 \frac{\nabla'_i \nabla'_j}{\nabla'^2}e_{jm}(\nabla'_i h' \nabla'_m h')\rightarrow  \lambda_2 \frac{\nabla_i \nabla_j}{\nabla^2}e_{jm}(\nabla_i h \nabla_m h)-\lambda_2e_{ij}c_i\frac{\partial h}{\partial x_j}.
\end{align}
Noise term remains invariant under this transformation. After the transformation, the extra terms in RHS are 
\begin{align}
\lambda \boldsymbol \nabla h \cdot \textbf{c}+\frac{\lambda}{2}c^2+2\lambda_1 \textbf{c}\cdot\boldsymbol\nabla h-\lambda_2e_{ij}c_i\frac{\partial h}{\partial x_j}.\label{gal_inv_1}
\end{align}
 Similarly LHS of Eq.~(\ref{ch-kpz}) transforms as,
\begin{align}
 \frac{\partial h'}{\partial t'}\rightarrow \frac{\partial h}{\partial t}+(\lambda+2\lambda_1)\textbf{c}\cdot\boldsymbol\nabla h-\lambda_2e_{ij}c_i\frac{\partial h}{\partial x_j}+(\lambda+2\lambda_1)c^2.\label{gal_inv_2}
\end{align} 
  Comparing Eq.~(\ref{gal_inv_1}) and Eq.~(\ref{gal_inv_2}) we see that Eq.~(\ref{ch-kpz}) is invariant.


\section{Renormalisation group calculations}

The dynamic renormalisation group calculation is conveniently performed in terms of a path integral over $h({\bf r},t)$ and its dynamic conjugate field $\hat h({\bf r},t)$~\cite{tauber,bausch} that is equivalent to and constructed from Eq.~(\ref{ch-kpz}) together with the noise variance. 
The generating functional corresponding to Eq.~(\ref{ch-kpz}) is given by~\cite{bausch,tauber}
\begin{equation}
 \mathcal{Z}=\int \mathcal{D}\hat{h} \mathcal{D}h e^{-\mathcal{S}[\hat{h},h]},
\end{equation}
where $\hat{h}$ is the dynamic conjugate field and $\mathcal{S}$ is the action functional:

{
\begin{align} 
S= -\int_{{\bf x},t}\hat{h}D\hat{h} + \int_{{\bf x},t}\hat{h}\bigg\{\partial_th-\nu \nabla^2h-\frac{\lambda}{2} (\boldsymbol{\nabla} h)^2-\lambda_1 \frac{\nabla_i \nabla_j}{\nabla^2}(\nabla_ih \nabla_jh)-\lambda_2 \frac{\nabla_i \nabla_j}{\nabla^2}e_{jm}(\nabla_i h \nabla_m h)\bigg\}. \label{action}
\end{align}
Action functional $S$ can be split into two parts. The first one is the Gaussian part $S_{1}$ and other one is the anharmonic part, containing the  local ($\lambda$ term) and  nonlocal ($\lambda_1$ and $\lambda_2$ terms) anharmonic terms,  $S_{2}$. Here,
\begin{align*}
&S_1=-\int_{{\bf k_1},\omega_1}\hat h(-{\bf k_1},-\omega_1)D\hat h({\bf k_1},\omega_1)+\int_{{\bf k_1},\omega_1}\hat h(-{\bf k_1},-\omega_1)\bigl\{ -i\omega_1+\nu k_1^2 \bigl\}h({\bf k_1},\omega_1)\\& S_2=\int_{{\bf k_2},\omega_2}\int_{{\bf k_3},\omega_3}\hat h(-{\bf k},-\omega)\Bigl\{ \frac{\lambda}{2}{\bf k_2\cdot k_3}+\lambda_1\frac{k_ik_j}{k^2}(k_{2i}k_{3j})+\lambda_2\frac{k_ik_j}{k^2}e_{jm}(k_{2i}k_{3m}) \Bigl\}h({\bf k_2},\omega_2)h({\bf k_3},\omega_3).
\end{align*}
We have defined ${\bf k=k_2+k_3}$ and $\omega=\omega_2+\omega_3$. A factor of $1/k^2$ in the  terms with coefficients $\lambda_1$ and $\lambda_2$  indicate their nonlocal nature in real space. This form of $S_2$ holds in 2D; at $d>2$, we must set $\lambda_2=0$, since the chiral term is 2D specific. As discussed in the main text, these nonlocal terms are contributions to the surface velocity that depend {\em nonlocally} on the local surface fluctuations.}


\subsection{Linear theory results}

We define Fourier transform by 
\begin{equation*}
a({\bf x},t)=\int_{{\bf q},\omega}a({\bf q},\omega)e^{i({\bf q\cdot x}-\omega t)},
\end{equation*}
where $a=h,\hat h$.
Then from the Gaussian part of (\ref{action}), the correlation functions at the harmonic order can be found:
\begin{subequations}
 \begin{align}
&\langle \hat{h}({\bf q},\omega) \hat{h}(-{\bf q},-\omega)\rangle_0=0 \\
&\langle \hat{h}({\bf q},\omega) h(-{\bf q},-\omega)\rangle_0=\frac{1}{i\omega +\nu q^2}\\
&\langle \hat{h}(-{\bf q},-\omega) h({\bf q},\omega)\rangle_0=\frac{1}{-i\omega +\nu q^2}\\
&\langle h({\bf q},\omega) h(-{\bf q},-\omega)\rangle_0=\frac{2D}{\omega^2 +\nu^2 q^4}
\end{align}\label{scaling}
\end{subequations}
\begin{figure}[]
\centering
\includegraphics[width=0.30\textwidth]{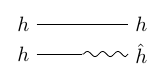}
\caption{Diagrammatic representation of two point functions.}
\label{propagator}
\end{figure}
Fig.~\ref{propagator} shows diagrammatic representation of the propagators.

\subsection{Corrections to $D$}
There are total four Feynman one-loop diagrams which contribute to the fluctuation-corrections of $D$.

\begin{figure}[htb]
\includegraphics[width=0.30\textwidth]{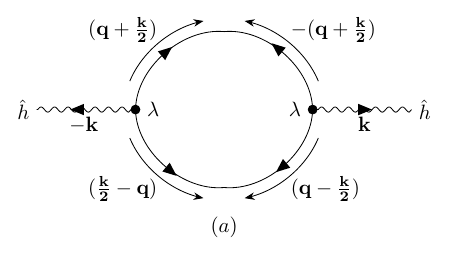}
\includegraphics[width=0.30\textwidth]{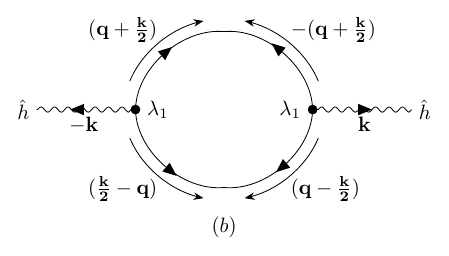}
\includegraphics[width=0.30\textwidth]{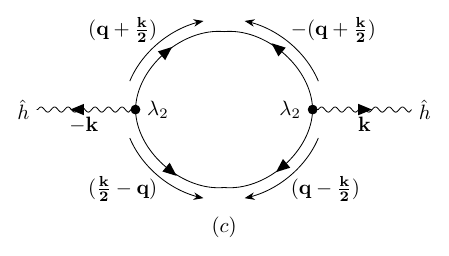}
\includegraphics[width=0.30\textwidth]{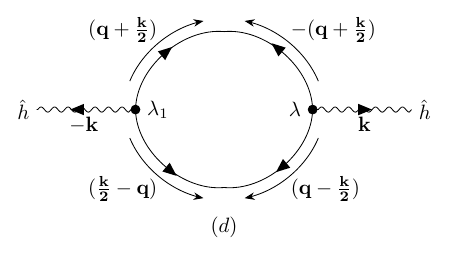}
\caption{ One-loop Feynman diagrams that contribute to the renormalisation of $D$}
\label{fig_1}
\end{figure}

Fig.~\ref{fig_1}$(a)$ has a symmetry factor $2$ and contributes to the correction of $D$ as shown below,

\begin{align}
 &\frac{2\lambda^2}{2!\times4}\int_{{\bf q},\Omega}\frac{\bigl({\bf (q+\frac{k}{2})}\cdot {\bf (\frac{k}{2}-q)}\bigl)^2\times 4D^2}{\Bigl(\Omega^2+\nu^2({\bf q+\frac{k}{2}})^4\Bigl)\Bigl(\Omega^2+\nu^2({\bf \frac{k}{2}-q})^4\Bigl)}\\&=\frac{\lambda^2D^2}{4\nu^3}\int\frac{d^2q}{q^2},
\end{align}

neglecting subleading contributions.
Fig.~\ref{fig_1}$(b)$ has a symmetry factor $2$ and contributes to the correction of $D$ as shown below,

\begin{align}
 &-\frac{\lambda_1^2}{2!}\times  2 \int_{{\bf q},\Omega}\frac{k_ik_j}{k^2}({\bf q+\frac{k}{2}})_i({\bf \frac{k}{2}-q})_j\frac{k_mk_n}{k^2}({\bf q+\frac{k}{2}})_m \times ({\bf q-\frac{k}{2}})_n  \times \frac{4D^2}{\Bigl(\Omega^2+\nu^2{\bf (q+\frac{k}{2})}^4\Bigl)\Bigl(\Omega^2+\nu^2{\bf (\frac{k}{2}-q)}^4\Bigl)}\\& =\frac{\lambda_1^2D^2}{\nu^3}\times\frac{3}{8}\int\frac{d^2q}{q^2},
\end{align}

again neglecting subleading contributions.
Fig.~\ref{fig_1}$(c)$ has a symmetry factor $2$ and contributes to the correction of $D$ as shown below,

\begin{align}
 &-\frac{\lambda_2^2}{2!}\times  2 \int_{{\bf q},\Omega}\frac{k_ik_j}{k^2}e_{jm}({\bf q+\frac{k}{2}})_i({\bf \frac{k}{2}-q})_m\frac{k_sk_t}{k^2}e_{tr}({\bf q+\frac{k}{2}})_s({\bf q-\frac{k}{2}})_r \times \frac{2D}{\Bigl(\Omega^2+\nu^2{\bf (q+\frac{k}{2})}^4\Bigl)} \times \frac{2D}{ \Bigl(\Omega^2+\nu^2{\bf (\frac{k}{2}-q)}^4\Bigl)}\\&=\frac{\lambda_2^2D^2}{8\nu^3}\int \frac{d^2q}{q^2},
 \end{align}

again neglecting subleading contributions.
 Fig.~\ref{fig_1}$(d)$ has a symmetry factor $2$ and contributes to the correction of $D$ as shown below,
 
 \begin{align}
 &-\frac{\lambda\lambda_1}{2}\times 2\frac{k_ik_j}{k^2} \int_{{\bf q},\Omega}({\bf q+\frac{k}{2}})_i({\bf \frac{k}{2}-q})_j\Bigl({\bf (q+\frac{k}{2})}\cdot {\bf (q-\frac{k}{2})}\Bigl) \times \frac{4D^2}{\Bigl(\Omega^2+\nu^2{\bf (q+\frac{k}{2})}^4\Bigl)\Bigl(\Omega^2+\nu^2{\bf (\frac{k}{2}-q)}^4\Bigl)}\\&=\frac{\lambda\lambda_1 D^2}{2\nu^3}\int \frac{d^2q}{q^2},
\end{align}

neglecting subleading contributions.
Adding all these contributions, we obtain the total corrections to $D$:
\begin{align}
D^{<}=D \Bigr[ 1+\Bigl(\frac{\lambda^2D}{4\nu^3}+\frac{3\lambda_1^2D}{8\nu^3}+\frac{\lambda_2^2D}{8\nu^3}+\frac{\lambda\lambda_1D}{2\nu^3}\Bigl)\int \frac{d^2q}{q^2} \Bigr]
\end{align}

\subsection{Corrections to $\nu$}

\begin{figure}[htb]
\includegraphics[width=0.30\textwidth]{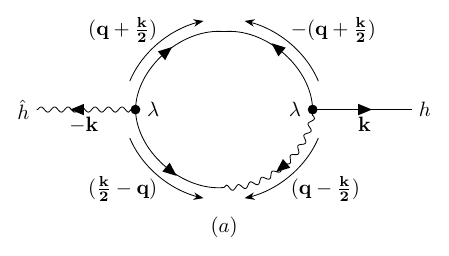}
\includegraphics[width=0.30\textwidth]{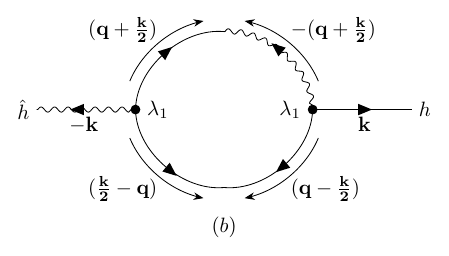}
\includegraphics[width=0.30\textwidth]{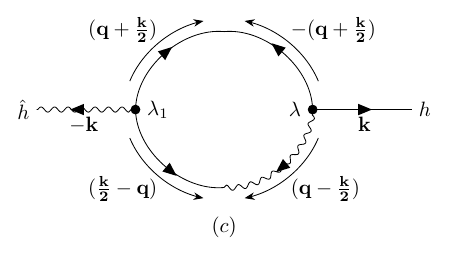}
\includegraphics[width=0.30\textwidth]{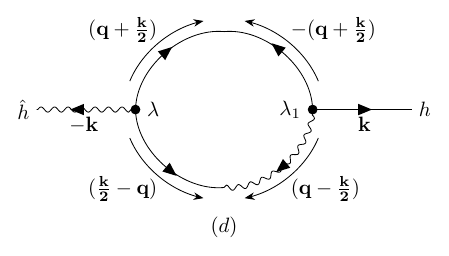}
\includegraphics[width=0.30\textwidth]{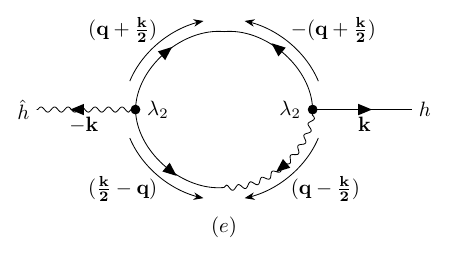}
\caption{ One-loop Feynman diagrams that contribute to the renormalisation of $\nu$}
\label{fig_2}
\end{figure}

Fig.~\ref{fig_2}$(a)$ represents the one-loop contribution proportional to $\lambda^2$, which has a symmetry factor of $8$ and is given by

\begin{align*}
-\frac{\lambda^2}{4}\times \frac{8}{2!}\int_{{\bf q},\Omega}\Bigl({\bf (q+\frac{k}{2})}\cdot {\bf (\frac{k}{2}-q)}\Bigl)\Bigl({\bf (q+\frac{k}{2})\cdot k}\Bigl)\times \frac{2D}{\Omega^2+\nu^2{\bf (q+\frac{k}{2})}^4} \times \frac{1}{i\Omega+\nu ({\bf \frac{k}{2}-q})^2}
\end{align*}

After performing the $\Omega$ integral above result becomes,
\begin{align*}
-\frac{\lambda^2 D}{\nu^2}\int_{\bf q}\frac{({\bf q+\frac{k}{2}})_i({\bf \frac{k}{2}-q})_i({\bf q+\frac{k}{2}})_jk_j}{({\bf q+\frac{k}{2}})^2\bigl\{{ \bf (q+\frac{k}{2})}^2+({\bf\frac{k}{2}-q)}^2\bigl\}}=\frac{\lambda^2D}{4\nu^2}\int_{\bf q}\frac{q^2q_jk_j+q^2k^2}{{ \bf (q+\frac{k}{2})}^2(q^2+\frac{k^2}{4})}.
\end{align*}
This then consists of two parts, coming from the two terms of the above integral.
Since correction to $\nu$ is of the order of $k^2$, the first term in the numerator of the above integral requires binomial expansion of the denominator. After expanding for small $k$, the first term of the above integral becomes
\begin{align*}
-\frac{\lambda^2D}{2\nu^2}k^2\times \frac{1}{d}\int\frac{d^dq}{q^2}.
\end{align*}
This may be evaluated by using the well-known relation~\cite{yakhot}
\begin{align} 
k_ik_j\int d^dqf(q^2)q_iq_j=k_ik_j\times\frac{[\delta_{ij}]}{d}\int d^dq f(q^2)q^2,\label{identity_1}
\end{align}
where $f(q^2)$ is a function of $q^2$.
 The second term of the above integral becomes contributes
\begin{align*}
\frac{\lambda^2D}{4\nu^2}k^2\times \int\frac{d^dq}{q^2}
\end{align*}
to the correction to $\nu$.
Adding above two results we get the total corrections to $\nu$ at ${\cal O}(\lambda)^2$. In 2D this takes following form
\begin{align}
\frac{\lambda^2D}{4\nu^2}\frac{(d-2)}{d}k^2\int \frac{d^2q}{q^2}
\end{align}

Fig.~\ref{fig_2}$(b)$ has symmetry factor $8$ and represents the one-loop contribution proportional to $\lambda_1^2$ to $\nu$. This reads

\begin{align*}
-\frac{\lambda_1^2}{2!}\times 8 \int_{{\bf q},\Omega} \frac{k_ik_j}{k^2}\Bigl(({\bf q+\frac{k}{2}})_i({\bf \frac{k}{2}-q})_j\Bigl)\frac{({\bf q+\frac{k}{2}})_m({\bf q+\frac{k}{2}})_n}{\bf (q+\frac{k}{2})^2}\times k_m({\bf \frac{k}{2}-q})_n \times \frac{2D}{\Omega^2+\nu^2({\bf \frac{k}{2}-q})^4}\times \frac{1}{-i\Omega + \nu ({\bf q+\frac{k}{2}})^2} 
\end{align*}

After doing the $\Omega$-integral above equation becomes,
\begin{align*}
-\frac{4\lambda_1^2D}{\nu^2} \frac{k_ik_jk_m}{k^2} \int_{\bf q} \frac{({\bf q+\frac{k}{2}})_i ({\bf \frac{k}{2}-q})_j ({\bf q+\frac{k}{2}})_m}{({\bf q+\frac{k}{2}})^2 ({\bf \frac{k}{2}-q})^2(2q^2+\frac{k^2}{2})}\times \frac{({\bf q+\frac{k}{2}})_n ({\bf \frac{k}{2}-q})_n}{({\bf q+\frac{k}{2}})^2 ({\bf \frac{k}{2}-q})^2(2q^2+\frac{k^2}{2})}
\end{align*}
Since we are interested in the corrections to $\nu k^2$, it suffices to retain the ${\cal O}(k^0)$ and ${\cal O}(k^1)$ parts in the numerator, giving
\begin{align*}
&\frac{4\lambda_1^2D}{\nu^2} \frac{k_ik_jk_m}{k^2} \int_{\bf q} \frac{(q_iq_m\frac{k_j}{2} -q_jq_m\frac{k_i}{2} -q_iq_jq_m -q_iq_j\frac{k_m}{2})q^2}{({\bf q+\frac{k}{2}})^2 ({\bf \frac{k}{2}-q})^2(2q^2+\frac{k^2}{2})}.
\end{align*}
First two terms in the numerator of the above integral vanish as, $k_ik_j$ is symmetric but $q_m(q_i\frac{k_j}{2}-q_j\frac{k_i}{2})$ is antisymmetric under the interchange of $i,j$. Third term of the integrand can be expanded for small $k$, which then takes the following form $-q_iq_jq_m({\bf k\cdot q}-{\bf k\cdot q})$ and vanishes identically. Only the fourth term contributes and can be evaluated using the identity~(\ref{identity_1}). Thus above integral reduces to

\begin{align}
-\frac{\lambda_1^2D}{2\nu^2}k^2\int \frac{d^2q}{q^2}.
\end{align}
Fig.~\ref{fig_2}$(c)$ has a symmetry factor $4$ and represents one of the contributions proportional to $\lambda_1\lambda$ where the external leg with $\hat{h}$ is from the vertex $\lambda_1$. This contribution is

\begin{align*}
-\frac{\lambda_1 \lambda}{2}\times 4 \times \frac{k_ik_j}{k^2}\int_{{\bf q},\Omega} ({\bf q+\frac{k}{2}})_i({\bf \frac{k}{2}-q})_j\Bigl({\bf (q+\frac{k}{2})\cdot k}\Bigl) \times \frac{2D}{\Omega^2+\nu^2({\bf q+\frac{k}{2}})^4}\times \frac{1}{i\Omega + \nu ({\bf \frac{k}{2}-q})^2} 
\end{align*}

After doing the $\Omega$ integral above integral becomes,
\begin{align*}
-\frac{2\lambda\lambda_1D}{\nu^2} \frac{k_ik_jk_m}{k^2}\int_{\bf q} \frac{({\bf q+\frac{k}{2}})_i({\bf \frac{k}{2}-q})_j({\bf q+\frac{k}{2}})_m}{({\bf q+\frac{k}{2}})^2(2q^2+\frac{k^2}{2})}=-\frac{2\lambda\lambda_1D}{\nu^2} \frac{k_ik_jk_m}{k^2}\int_{\bf q} \frac{q_i(q_m\frac{k_j}{2}-q_j\frac{k_m}{2})-\frac{k_i}{2}q_jq_m-q_iq_jq_m}{({\bf q+\frac{k}{2}})^2(2q^2+\frac{k^2}{2})}
\end{align*}
The first term in the numerator of the above integral vanish as $k_jk_m$ is symmetric but $q_i(q_m\frac{k_j}{2}-q_j\frac{k_m}{2})$ is antisymmetric under the interchange of $j,m$. We evaluate the third term by expanding for small $k$
\begin{align*}
\frac{2\lambda\lambda_1D}{\nu^2} \frac{k_ik_jk_m}{k^2}\Biggr[\int_{\bf q} \frac{\frac{k_i}{2}q_jq_m}{2q^4}-\int_{\bf q} \frac{q_iq_jq_mq_nk_n}{2q^6}\Biggr].
\end{align*}
The first integral in the above line can be calculated using identity~(\ref{identity_1}). The second integral can be done by using the following identity~\cite{yakhot}
\begin{align} 
k_ik_jk_mk_n\int d^dqf(q^2)q_iq_jq_mq_n=k_ik_jk_mk_n\times\frac{[\delta_{ij}\delta_{mn}+\delta_{im}\delta_{jn}+\delta_{in}\delta_{jm}]}{d(d+2)}\int d^dq f(q^2)q^4\label{identity_2}.
\end{align}
Thus calculating the above integrals and substituting $d=2$ we get,

\begin{align}
-\frac{\lambda \lambda_1D}{8\nu^2}k^2\int \frac{d^2q}{q^2}
\end{align}
Fig.~\ref{fig_2}$(d)$ has symmetry factor $4$ and represents the contribution originating from vertices $\lambda_1$ - $\lambda$ where external leg $\hat{h}$ is from vertex $\lambda$. The contribution is,

\begin{align}
&-\frac{\lambda \lambda_1}{2}\times 4 \int_{{\bf q},\Omega} \Bigl({\bf (q+\frac{k}{2})}\cdot {\bf (\frac{k}{2}-q)}\Bigl)\frac{({\bf \frac{k}{2}-q})_i({\bf \frac{k}{2}-q})_j}{{\bf(\frac{k}{2}-q)}^2}\times ({\bf q+\frac{k}{2}})_ik_j \times \frac{2D}{\Omega^2+\nu^2({\bf q+\frac{k}{2}})^4}\times \frac{1}{i\Omega + \nu ({\bf \frac{k}{2}-q})^2}\\&=-\frac{\lambda \lambda_1D}{2\nu^2}k^2\int \frac{d^2q}{q^2}.
\end{align}

Fig.~\ref{fig_2}$(e)$ has a symmetry factor $2$ and represents the contribution proportional to $\lambda_2^2$. It is given by

\begin{align}
&-\frac{\lambda_2^2k_ik_j}{k^2}\int_{{\bf q},\Omega} \Bigl(({\bf q+\frac{k}{2}})_i({\bf \frac{k}{2}-q})_m+({\bf \frac{k}{2}-q})_i({\bf q+\frac{k}{2}})_m\Bigl)\times e_{jm}e_{tn} \frac{({\bf \frac{k}{2}-q})_s({\bf \frac{k}{2}-q})_t}{({\bf \frac{k}{2}-q})^2}\Bigl(({\bf q+\frac{k}{2}})_sk_n+({\bf q+\frac{k}{2}})_nk_s\Bigl)\nonumber\\& \times \frac{2D}{\Omega^2+\nu^2({\bf q+\frac{k}{2}})^4}\times \frac{1}{i\Omega + \nu ({\bf \frac{k}{2}-q})^2}\\&=-\frac{\lambda_2^2D}{8\nu^2}k^2\int \frac{d^2q}{q^2}
\end{align}

There are no corrections proportional to $\lambda\lambda_2$ or $\lambda_1\lambda_2$ due to the chiral form of the $\lambda_2$-term.
Adding all these we find the total corrections to $\nu$:
\begin{align}
\nu^{<}=\nu \Bigr[ 1+\Bigl(\frac{\lambda_1^2D}{2\nu^3}+\frac{5}{8}\frac{\lambda\lambda_1D}{\nu^3}+\frac{\lambda_2^2D}{8\nu^3}\Bigl)\int \frac{d^2q}{q^2} \Bigr]
\end{align}\

\subsection{One-loop corrections to $\lambda,\lambda_1$ and $\lambda_2$}

\begin {figure}[htb]
\includegraphics[width=0.238\textwidth]{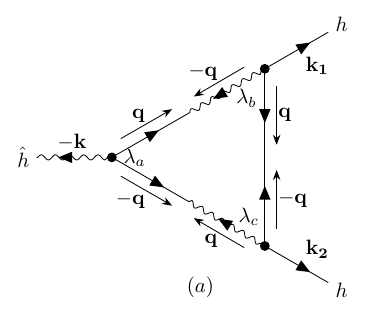}
\includegraphics[width=0.238\textwidth]{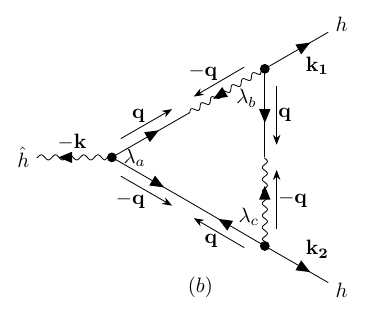}
\caption{ Representative one-loop Feynman diagrams that can contribute to the renormalisation of $\lambda$, $\lambda_1$ and $\lambda_2$. }
\label{fig_3}
\end {figure}
Fig.~\ref{fig_3} shows representative one-loop Feynman diagrams for renormalisation of $\lambda$, $\lambda_1$ and $\lambda_2$ at the one-loop order, where $a,b,c$ can take values $0,1,2$, and $\lambda_0=\lambda$.
There are many possible similar diagrams depending on $a,b,c$ but for a particular set of vertices (i.e for a fixed values of $a,b,c$) there exist two diagrams that may contribute to the corrections of any of $\lambda,\lambda_1,\lambda_2$ as shown in Fig.~\ref{fig_3}. When ($a=b=c=1$) then Fig.~\ref{fig_3}$(a)$ has symmetry factor $24$ and contributes following, 
\begin{align}
&\frac{\lambda_1^3}{3!}\times 24\times \Bigr[\frac{\nabla_i\nabla_j}{\nabla^2}\hat h \nabla_mh\nabla_th\Bigr]\int_{{\bf q},\Omega} q_iq_jq_mq_t \times \frac{2D}{(\Omega+i\nu q^2)^2(\Omega-i\nu q^2)^2} \\&=\frac{48\times \lambda_1^3}{3!\times8} \times \Bigr[\frac{\nabla_i\nabla_j}{\nabla^2}\hat h \nabla_mh\nabla_th\Bigr]\frac{2D}{8\nu^3}\int \frac{d^2q}{q^2} \times \Bigr[\delta_{ij}\delta_{mt}+\delta_{im}\delta_{jt}+\delta_{it}\delta_{jm}\Bigr]\label{lam1}
\end{align}
Similarly for ($a=b=c=1$) Fig.~\ref{fig_3}$(b)$  has symmetry factor $48$ and contributes following,

\begin{align}
&\frac{48\lambda_1^3}{3!}\times \Bigr[\frac{\nabla_i\nabla_j}{\nabla^2}\hat h \nabla_mh\nabla_th\Bigr]\int_{{\bf q},\Omega} q_iq_jq_mq_t\times \frac{2D}{(\Omega^2+\nu^2q^4)(\Omega+i\nu q^2)^2} \\
&=-\frac{48\times \lambda_1^3}{3!\times8} \times \Bigr[\frac{\nabla_i\nabla_j}{\nabla^2}\hat h \nabla_mh\nabla_th\Bigr]\frac{2D}{8\nu^3}\int \frac{d^2q}{q^2} \times \Bigr[\delta_{ij}\delta_{mt}+\delta_{im}\delta_{jt}+\delta_{it}\delta_{jm}\Bigr]\label{lam2}
\end{align} 
We can see that Eq. (\ref{lam2}) is exactly same as Eq. (\ref{lam1}) but with a negative sign. They cancel each other contributing nothing to the corrections of $\lambda,\lambda_1$ and $\lambda_2$.
In fact, for any combination of $a,b$ and $c$ there exist diagrams like Fig.~\ref{fig_3}$(a)$ and Fig.~\ref{fig_3}$(b)$ added up to zero contributing nothing to $\lambda,\lambda_1$ and $\lambda_2$.. This shows that there are no relevant corrections to $\lambda,\lambda_1$ and $\lambda_2$ at the one-loop order.

\subsection{RG flow equations and scaling}

After averaging over fields with higher momentum new action contains fields with upper cut off momentum $\frac{\Lambda}{b}$. Since we want to describe the same system with new action, so we have to rescale the space, time and fields so the upper cutoff momentum becomes $\Lambda$.
Rescaling of space and time give ${\bf q} \to {\bf q}'=b{\bf q} \implies x'=\frac{x}{b}$ and $\omega \to \omega'=b^z\omega \implies t'=\frac{t}{b^z}$. 
Furthermore, rescaling of the fields $h$ and $\hat h$ give
\begin{align*}
    &h^<({\bf q},\omega)=\xi h({\bf q}',\omega')\\
     &\hat h^<({\bf q},\omega)=\hat \xi \hat h({\bf q}',\omega')\\
     &h^<(x,t)=\xi_R h(x',t').
\end{align*}
Now let $\xi_R=b^\chi$, where $\chi$ is the roughness exponent of the surface. Since there is no correction to $\int_{x,t}\hat{h}\partial_t\,h$ we impose the condition that coefficient of $\int_{x,t}\hat{h}\partial_th$ remains unity under rescaling. Using these we find 
\begin{align*}
&D'=D^<b^{z-d-2\chi}\\
   &\nu'=\nu^<b^{z-2}\\
   &\lambda'=\lambda b^{ \chi+z-2}.
\end{align*}
Here, $\lambda_1$ and $\lambda_2$ scale same as $\lambda$. Since there are no relevant one-loop corrections to $\lambda,\lambda_1,\lambda_2$, $\lambda^<=\lambda$ etc. Furthermore,
$$D^{<}=D \Bigr[ 1+\Bigl(\frac{\lambda^2D}{4\nu^3}+\frac{3\lambda_1^2D}{8\nu^3}+\frac{\lambda_2^2D}{8\nu^3}+\frac{\lambda\lambda_1D}{2\nu^3}\Bigl)\int_{\frac{\Lambda}{b}}^{\Lambda} \frac{d^2q}{q^2} \Bigr]$$ and 
$$\nu^{<}=\nu \Bigr[ 1+\Bigl(\frac{\lambda_1^2D}{2\nu^3}+\frac{5}{8}\frac{\lambda\lambda_1D}{\nu^3}+\frac{\lambda_2^2D}{8\nu^3}\Bigl)\int_{\frac{\Lambda}{b}}^{\Lambda} \frac{d^2q}{q^2} \Bigr].$$ We set $b=e^{\delta l}\approx 1+\delta l$ and define dimensionless coupling constants by $g=\frac{\lambda^2D}{\nu^3}\frac{1}{2\pi}$,$\gamma_1 = \frac{\lambda_1}{\lambda}$ and $\gamma_2 = \frac{\lambda_2}{\lambda}$ to obtain the RG flow equations in the main text.

In dimensions greater than 2 (where $\gamma_2=0$), we set $d = 2 + \epsilon$ with $\epsilon > 0$. We aim to determine the fixed point and the corresponding scaling exponents accurately up to $O(\epsilon)$ only. To achieve this, it suffices to substitute $d = 2$ in the one-loop integrals. For example, in a general dimension $d>2$, the RG flow equation governing the parameter $g$ is expressed as follows:
%
\begin{align*}
\frac{dg}{dl}=g\Biggr[2-d+g\biggl\{ \frac{1}{4}+\frac{3}{d(d+2)}\gamma_1^2+\frac{\gamma_1}{d}-\frac{3}{4}\frac{2-d}{d}-\frac{3}{d}\gamma_1^2-\frac{3\gamma_1}{d(d+2)}\biggl(3-\frac{d+2}{2}+\frac{d(d+2)}{2}\biggl) \biggl\}\Biggr]
\end{align*} 
Upon setting $d=2+\epsilon$, and considering that $g\sim {\cal O}(\epsilon)$ at the fixed point, we substitute $d=2$ selectively in the contributions coming from the one-loop integral terms containing $d$ inside the curly bracket. This yields the simplified form of the RG flow equation:
\begin{align*}
 \frac{dg}{dl}=-\epsilon g - {\cal \tilde A}(\gamma_1) g^2,
\end{align*}
that is correct to ${\cal O}(\epsilon)$.
Consequently, this produces $g$ correctly to ${\cal O}(\epsilon)$; see, e.g., Ref.~\cite{stanley}.\\ 

\section{Correlation function}

 Following Refs.~\cite{activexy1,activexy2,sm1}, we now calculate the renormalised correlation functions of $h({\bf x})$, defined as
\begin{equation}
 C_{h}^R( r)\equiv \langle [h({\bf x})-h({\bf x'})]^2\rangle_R,
\end{equation}
where $R$ refers to a renormalised quantity.
We start from
\begin{equation}
  \langle h({\bf k})h({\bf -k})\rangle_R \approx \frac{ D_0}{\nu_0 k^2[\ln (\Lambda/k)]^{1-\mu}},\label{anhar-k}
\end{equation}
Expression (\ref{anhar-k}) is no longer valid over the wavevector range from 0 to $\Lambda$, rather it is valid between 0 and $\tilde\Lambda\ll \Lambda$.
 We then obtain,
 \begin{equation}
  C_{h}(r)\approx\int_0^{\tilde\Lambda} \frac{d^2k}{(2\pi)^2}\left[1- \exp i{\bf k}\cdot ({\bf x-x}')\right]\frac{ D_0}{ \nu_0 k^2 [\ln (\Lambda/k)]^{1-\mu}}.\label{anhar-r}
\end{equation}
Integrating over the angular variable, we get
\begin{eqnarray}
 C_{h}(r)&\approx&\int_0^{\tilde\Lambda}\frac{dq\,D_0}{\nu_0 q[\ln(\Lambda/q)]^{1-\mu}}\left[\frac{1}{2\pi} \int_0^{2\pi} d\theta (1-e^{iqr\cos\theta}\right]\nonumber \\
 &=& \int_0^{\tilde\Lambda}\frac{dq\, D_0}{\nu_0q[\ln(\Lambda/q)]^{1-\mu}}\left[1-J_0(qr)\right]\nonumber \\
 &=& \int_0^{\tilde\Lambda r}\frac{du\, D_0[1-J_0(u)]}{\nu_0 u |\ln (\frac{r\Lambda}{u})|^{1-\mu}},
 \end{eqnarray}
where $J_0(u)$ is the Bessel function of order zero. Then
 \begin{equation}
  C_{h}(r)=\int_0^1\frac{du\,D_0[1-J_0(u)]}{\nu_0u[-\ln u + \ln (1/y)]^{1-\mu}}+\int_1^{\tilde\Lambda r}\frac{du\, D_0}{\nu_0u[-\ln u + \ln (\Lambda r)]^{1-\mu}} - \int_1^{\tilde\Lambda r} \frac{du\, D_0J_0(u)}{\nu_0u[-\ln u + \ln (\Lambda r)]^{1-\mu}}. \label{inter-appex}
 \end{equation}
 The first and the third terms on the rhs of (\ref{inter-appex}) are finite. Since $u_{max}=\tilde\Lambda r \ll \Lambda r$, the second contribution on the right may be integrated with the substitution $u=\exp(z)$ giving
 \begin{equation}
  \int_1^{\tilde\Lambda\, r}\frac{du}{u[\ln (\Lambda\,r)]^{1-\mu}} 
  =\mu \frac{\ln (\Lambda r)+\ln (\tilde\Lambda/\Lambda)}{\ln (\Lambda r)}[\ln(\Lambda r)]^\mu\approx \mu[\ln(\Lambda\,r)]^{\mu}
 \end{equation}
 
 in the limit of large $r$.
 We thus find $C_{h}(r)\approx \mu\frac{D_0}{\nu_0}|\ln(\Lambda\,r)|^{\mu}$ in the limit of large $r$, with the remaining contributions on the right hand side of (\ref{inter-appex}) being finite or subleading for large $r$.
\end{document}